\documentclass[12pt]{iopart}
\usepackage[latin1]{inputenc}
\usepackage{iopams}
\usepackage[dvips]{graphicx}
\usepackage{epic,epsfig}
\usepackage{epstopdf}
\usepackage{subfigure}

\usepackage{dcolumn}
\usepackage{tabularx}
\begin{document}

\title{Detecting modules in dense weighted networks with the Potts method}

\author{Tapio Heimo$^{1,2}$}
\author{Jussi M. Kumpula$^{1}$}
\author{Kimmo Kaski$^{1}$}
\author{Jari Saram\"aki$^{1}$}

\address{$^{1}$ Department of Biomedical Engineering and Computational Science, Helsinki
  University of Technology, P.O. Box 9203, FIN-02015 HUT, Finland}
\address{$^{2}$ Nordea Bank AB, Markets Division, H224, SE-10571 Stockholm, Sweden}

\ead{taheimo@cc.hut.fi}

\begin{abstract}
We address the problem of multiresolution module detection in dense weighted networks, where the modular structure is encoded in the weights rather than topology. We discuss a weighted version of the $q$-state Potts method, which was originally introduced by Reichardt and Bornholdt. This weighted method can be directly applied to dense networks. We discuss the dependence of the resolution of the method on its tuning parameter and network properties, using sparse and dense weighted networks with built-in modules as example cases. Finally, we apply the method to stock price correlation data, and show that the resulting modules correspond well to known structural properties of this correlation network.
\end{abstract}

\maketitle

\newcommand{\ave}[1]{\langle #1 \rangle}
\newcommand{\h}{\mathcal{H}}

\section{Introduction}

During the recent years, the network approach has proven to be a very efficient way for investigating a wide range of complex systems \cite{RefWorks:133,RefWorks:178,RefWorks:73,RefWorks:179}. In this approach, the fundamental elements of the system are represented with nodes and the interactions between them with links. Sometimes it is enough to consider links as "binary", such that each link either exists or not. In this case, it is assumed that the pure topology carries enough relevant information about the system
under study. However, valuable information is often lost if interaction strengths are 
not taken into account. Because of this, the study of \emph{weighted} networks has recently been receiving a lot of attention. In this framework, a scalar weight representing the associated interaction strength is assigned to each link. It is evident that this additional degree of freedom somewhat  complicates the picture, for example generalization of existing measures is not necessarily straightforward (see, \emph{e.g.}, \cite{Saramaki:2007zr}). Thus there is a need for developing new network analysis methods which focus on the weights instead of pure topology.

The study of (weighted) networks has mostly focused on systems whose interaction structure is inherently sparse, such as air transport networks~\cite{Barrat:2004ye,Colizza:2006oa} or social networks inferred from electronic communication records~\cite{Onnela:2007fk,Lambiotte2008}. Another approach is to filter out interactions which are considered insignificantly weak, resulting in sparse network representations even for systems where each element interacts with each other, \emph{i.e.}, systems whose "natural" representation is a full or dense weighted network. For such networks, it is the interaction strengths themselves that carry the most significant information -- the networks are constructed on the basis of the assumption that the strongest interactions encode the most significant properties for the system under study. 
This is the case for instance with correlation-based networks, in which the weights are usually related to correlations between the time series of some relevant activities of the nodes (see, \emph{e.g.}, \cite{RefWorks:190}), or distance-based networks~\cite{Rozenfeld:2007uq}, in which the weights are related to distances between the nodes according to some relevant metric. It is evident that in this approach setting the proper threshold below which interactions are discarded is a non-trivial task.

In addition to weighted networks, the attention of network science has recently been focusing on "mesoscopic" properties of networks, \textit{i.e.}, structures beyond the scale of single nodes or their immediate neighborhoods. A very important and related problem is the detection and study of \textit{modules} or \textit{communities}\footnote{In this paper, these two terms will be used interchangeably.}, \textit{i.e.}, groups of nodes with dense internal connections and sparse connections to the rest of the network \cite{RefWorks:46,RefWorks:52,RefWorks:67,RefWorks:34,RefWorks:112,SantoReview}. A number of methods have been introduced, mostly in the context of binary networks. These include various modularity optimization methods building on the work by Newman and Girvan \cite{RefWorks:46}, the clique percolation method by Palla \textit{et al.} \cite{RefWorks:52}, and methods based
on statistical inference \cite{Hastings:2006fk,Hofman:2008uq}. Many methods have been generalized to deal with weighted networks \cite{RefWorks:105,RefWorks:204,RefWorks:173,RefWorks:142}; however, e.g. for the clique percolation method, networks have to be fairly sparse in order for the method to be applicable. Regarding the modularity optimization family of methods, it has been shown that there is an intrinsic resolution limit \cite{RefWorks:105,RefWorks:140,RefWorks:180}. However, a lot of attention has recently been given to \emph{multiresolution} methods \cite{RefWorks:34,RefWorks:204,RefWorks:142,RefWorks:180,RefWorks:50}, which allow investigating modular structure at various levels of coarse-graining. 

In this work we concentrate on investigating modular structure in dense weighted networks, using a weighted version of the $q$-state Potts method by Reichardt and Bornholdt (RB) \cite{RefWorks:34,RefWorks:50}. This method is closely related to modularity optimization methods, and hence
there is a resolution limit \cite{RefWorks:105}. However, the method contains a tuning parameter
which allows changing this limit. Although the method was originally introduced in the context of sparse, binary networks,
edge weights can readily be taken into account~\cite{RefWorks:50}. In fact, once this is done,
the networks to be analyzed need no longer to be sparse -- hence, for example when studying 
stock market correlations, all correlation matrix elements can be taken into account and no thresholding is necessary. 

 We begin by discussing the weighted RB method, deriving the required weighted null model, and then investigate the effect of the tuning parameter on the resolution of the method for networks with modular structure encoded in the weights. Then, we apply the method to a correlation-based network of stock return time series, \emph{i.e.}, a full correlation matrix, whose modular structure has been earlier investigated using a wide variety of approaches (see, \textit{e.g.}, \cite{RefWorks:190,RefWorks:183,RefWorks:184,RefWorks:192}).  It should be noted here that the multiresolution method recently introduced by Arenas \emph{et al.}~\cite{RefWorks:204} bears some similarity with the Potts method (see~\cite{RefWorks:180}); thus for comparison we apply it to the same data. Finally, we draw conclusions.

\section{The RB method}

\subsection{Introduction}

Let us begin with a short introduction of the community detection method introduced by
Reichardt and Bornholdt (RB)~\cite{RefWorks:34,RefWorks:50}. In this method, each node is assigned to exactly one module, and
the module indices of nodes
are considered as spins of a $q$-state Potts model. The goal is to assign 
nodes to modules in such a way that the energy of the system
is minimized. In the global optimum, groups of nodes with dense internal 
connections should end up having parallel spins. 
The Hamiltonian for the system is defined as:
\begin{equation}
\h_u = - \sum_{m} (l_{mm}-\gamma[l_{mm}]_{p_{ij}}),
\label{eq:RB1}
\end{equation}
where $l_{mm}$ is the number of links inside module $m$, $[l_{mm}]_{p_{ij}}$ is
the expected number of links inside module $m$ given the null model $p_{ij}$, and
$\gamma > 0$ is an adjustable parameter. The summation is over all modules. 
The null model $p_{ij}$ denotes the probability that
a link would exist between nodes $i$ and $j$ if the network was entirely random, \emph{i.e}, in the absence of modular structure. Essentially, 
there are two possible choices for the null model: constant $p_{ij}=p$, which corresponds to 
Erd\"os-Renyi networks \cite{RefWorks:71}, and the configuration model \cite{RefWorks:73}, in which
the degree sequence of the original network is retained but all links are randomly rewired, such that all correlations are lost to the extent allowed by the degree sequence.

Next we briefly review the derivation of 
$[l_{mm}]_{p_{ij}}$ for the
configuration model. Imagine that all the links in the network are cut in half, such that
nodes have stubs (\emph{i.e.}, half-links) connected to them. Then these stubs are to be 
randomly reconnected to form full links. If two such stubs are randomly picked,
the probability that both connect to nodes in module $m$ is simply $K_{m}^{2}/K^{2}$, where
$K$ is the degree sum of the network\footnote{The degree sum of the network is defined by $K=\sum_{i=1}^N k_i$, where $k_i$ is the degree of node $i$.} and $K_{m}$ the degree sum of nodes in module $m$. Since there are $K/2$ pairs of stubs, we get
\begin{equation}
[l_{mm}] = \frac{K_m^2}{2K}.
\label{eq:RB3}
\end{equation}
Correspondingly, the probability that the two stubs to be connected belong to different modules, say $m$ and $n$, is $2 K_{m}K_{n}/K^2$. Thus, the expected number of links between modules $m$ and $n$ reads
\begin{equation}
[l_{mn}] = \frac{K_m K_n}{K}.
\label{eq:RB2}
\end{equation}

Let us now address the question of weighted networks. It seems natural that equation (\ref{eq:RB1})
transforms to
\begin{equation}
\h_w = - \sum_{m} (w_{mm}-\gamma[w_{mm}]_{p_{ij}}),
\label{eq:FM1}
\end{equation}
where $w_{mm}$ and $[w_{mm}]_{p_{ij}}$ denote the sum of weights and expected sum of weights of links
inside module $m$, respectively. Again, there are essentially two ways to define $[w_{mm}]_{p_{ij}}$. The approach 
taken in \cite{RefWorks:105} is to calculate the expected number of links using the configuration model and to assume that each link has average weight, that is, $[w_{mm}] = \ave w [l_{mm}]$. However, here we take  another approach,
which is analogous to the above derivation for the unweighted case and based on the ideas presented in \cite{RefWorks:63}. In weighted networks, the \emph{strength} $s_i$ of node $i$ is defined as the sum of the weights of the links attached to it. Consider dividing the strength of each node in small "stubs" of weight $ds$ such that node $i$ has $s_{i}/ds$ stubs emerging from it and start randomly connecting pairs of these stubs. This process is analogous to the above unweighted case, and as a result the expected sums of weights of the links inside module $m$ and between modules $m$ and $n$ are
\begin{equation}
[w_{mm}] = \frac{S_m^2}{2S}, \quad \textrm{and} \quad  [w_{mn}] = \frac{S_m S_n}{S}, 
\label{eq:FM2}
\end{equation}
respectively, where $S=\sum_{i=1}^N s_i$ is the strength sum of the network and $S_{q}$ the strength sum of module $q$.
When all links have weight $w_{ij}=1$, the above equations reduce to equations (\ref{eq:RB3}) and (\ref{eq:RB2}).

\subsection{Resolution of the weighted RB method for sparse and dense networks}
\label{sec:limit}

\begin{figure}[!h]
\begin{center}
\includegraphics[width=0.6\textwidth]{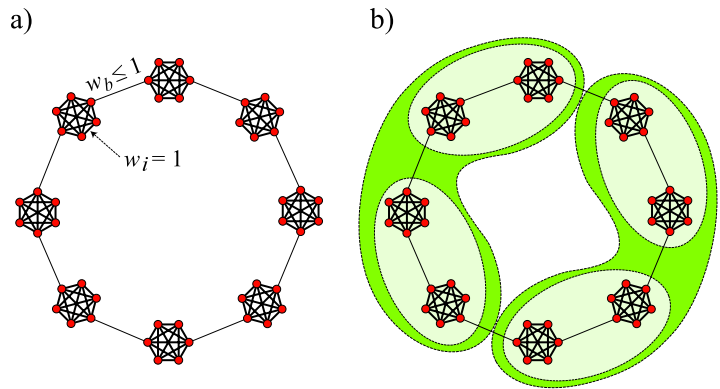}
\caption{a) A ring-like network, consisting of $N_b$ cliques, each containing of $N_c$ nodes. Link weights $w_i$ within modules equal unity, whereas modules are joined by links of weight $w_b\leq 1$. b) The weighted RB method can merge consecutive cliques to larger modules,
depending on values of the network parameters and the tuning parameter $\gamma$. The hierarchical structure is for illustrative purposes only. In general, the RB method does not yield hierarchical modules.}
\label{fig:ringnet}
\end{center}
\end{figure}

The RB method can be viewed as a general framework for community detection~\cite{RefWorks:50}, which for the unweighted case includes the modularity optimization method as a special case ($\gamma=1$ and configuration model as the null model). Recently, it was shown that the resolution of modularity optimization methods is intrinsically
 limited~\cite{RefWorks:140}. In particular,
in large networks small "physical" communities cannot be resolved and thus there is a lower limit to the size of communities which can be detected by the method. This limit depends on the number of links in the network and is also inherited by the more general RB method~\cite{RefWorks:105}. However, by changing the parameter $\gamma$, the resolution of the method can be tuned such that small values yield large modules and vice versa. This provides a clear advantage over "traditional" modularity optimization, which is restricted to a single resolution.

We now address the issue of resolution of the weighted RB method, beginning with a weighted modular network which is sparse, that is, whose average degree $\ave k \ll N$. Consider a simple case, where the $N$ nodes are arranged into modules of constant size $N_c$, so that the number of such modules is $N_b=N/N_c$. Let the modules form a ring-like structure, as illustrated in Fig.~\ref{fig:ringnet}, and let each module be a fully connected clique. Let the internal links within cliques have weight $w_i=1$,  and successive modules be connected by a single link of weight $w_b$, where $w_b \le 1$. This presents perhaps the simplest possible modular structure for a weighted connected network. 

The community structure
 found by the weighted RB method corresponds to the global minimum of the Hamiltonian (or energy) defined in Eq.~(\ref{eq:FM1}). Depending of the network parameters $N_b$, $N_c$, and $w_b$ as well as the tuning parameter $\gamma$, this structure may or may not correspond to the built-in modules. Let us consider two ways to group the built-in modules into communities: the first one is the ''natural'' grouping in which each built-in module is identified as a single community. In the second case, we take two successive built-in modules and consider them merged, that is, identified as one community. Other built-in modules are still considered as separate communities exactly as in the first case. Clearly, if the second grouping has smaller energy (\ref{eq:FM1}) than the first one, the resolution of the method is limited. 
A straightforward calculation shows that this is equal to the requirement
\begin{equation}
w_{mn} > \gamma [w_{mn}]=\gamma \frac {S_m S_n}{S}
\label{eq:sparse1}
\end{equation}
where $m$ and $n$ are the built-in modules to be merged, $S=\sum_{i=1}^N s_i$ is again the strength sum of the network, and $S_{q}$ the strength sum of module $q$.
 Now, $w_{mn}=w_b$, 
$S_m=S_n=N_c(N_c-1) + 2 w_b$, and $S=N_bS_m$. Plugging these into 
Eq. (\ref{eq:sparse1}) yields the merging condition for the example network:
\begin{equation}
w_b > \gamma \frac 1 {N_b} (N_c^2-N_c+2 w_b).
\label{eq:sparse2}
\end{equation}

Now, let the network size $N$ increase while the module size $N_c$ remains constant. Then, as $N_b=N/N_c$ increases, larger and larger values of $\gamma$ are needed for obtaining the built-in modules. Increasing $w_b$ makes merging easier, as expected. For $w_b=1$, Eq.~(\ref{eq:sparse2}) yields the resolution limit for the similar unweighted network studied in \cite{RefWorks:105}.

\begin{figure}[!h]
\begin{center}
\includegraphics[width=0.8\textwidth]{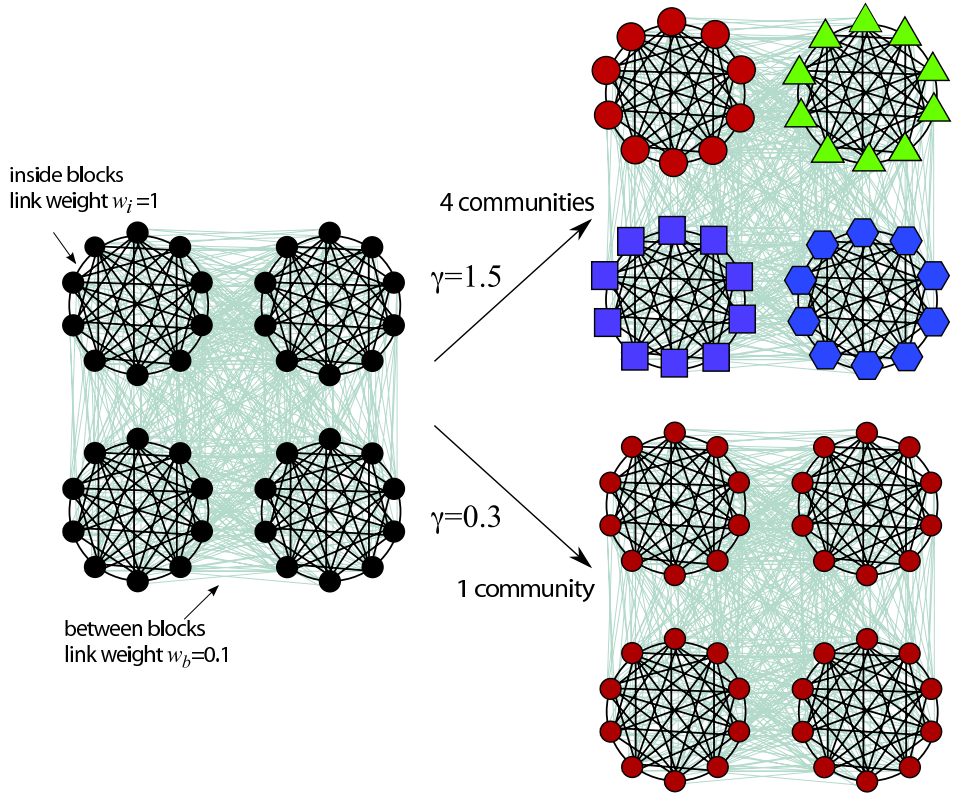}
\caption{Left: A network consisting of $N_b=4$ blocks each having $N_c=10$ nodes. Links inside blocks have weight $w_i=1$ and nodes in different 
blocks are connected with links of weight $w_b=0.1$. On the right is illustrated the effect of $\gamma$ on the found modular structure. Large values
yield the physical communities while for small values the communities appear as one large module. If the number of blocks $N_b$ is large enough, the networks size does not affect the $\gamma$ values where
merging happens. }
\label{fig:esim}
\end{center}
\end{figure}

Let us now move on to a more interesting case where the network in question is fully connected, \emph{i.e.}, links exist between each node, and the modular structure is purely encoded in the weights. 
Perhaps the simplest possible structure for a fully connected network with modules is the
case where $N_b$ modules each consisting of $N_c$ nodes are constructed such that
inside the modules the links have weight $w_i=1$ and links between nodes in different modules
have weight $w_b$ ($0 < w_b \leq 1$), see Fig.~\ref{fig:esim}. 
Similarly to the above analysis for the sparse
weighted network, we again consider two ways to group the built-in modules to communities:
the ''natural'' grouping and the one in which two built-in modules are considered as a single module.
Again, the method prefers the second grouping over the natural one if it yields smaller energy (Eq. (\ref{eq:FM1})).
The condition for this is again given by Eq. (\ref{eq:sparse1}),but now we have
$ w_{mn}  =  N_c^2 w_b $ and $ S_{q} = N_c s_i,$ 
where $s_i = N_c-1 + (N_b-1)N_c w_b $ denotes the (constant) strength of the nodes.
Thus, Eq. (\ref{eq:sparse1}) is equivalent to
\begin{equation}
N_c^2 w_b >   \gamma N_c^2 \left[ \frac{1-\frac 1 {N_c} }{N_b} + (1-\frac 1 {N_b})w_b \right]  \approx  \gamma N_c^2 w_b,
\label{eq:res2}
\end{equation}
where the approximation is valid when $N_b$ is large. In this case, Eq. (\ref{eq:res2}) further simplifies to $\gamma < 1$, where
it should be understood that the specific merging value $\gamma=1$ appears as a result of the simple structure of the
example case. In a more general scope,  
the expected weight between modules $[w_{mn}] \approx N_c^2 w_b$ is independent of the number
of modules $N_b$, i.e., network size. Thus, merging is solely controlled by $\gamma$. 
This is different from the sparse network case discussed above, where increasing
system size eventually triggers merging as the expected number and the total weight of links between modules decreases.

Finally, we analyse the effects of a single strong link between the modules in the latter example case.
On the basis of the above analysis, merging happens if the total weight between
the two modules exceeds $\gamma \left[w_{mn}\right]$, which is again of the order of $N_c^2w_b$. 
For sufficiently large $N_c$, the expected weight is so large that adding one strong link is
not enough for merging to occur. Smaller modules are merged more easily. However,
the resolution limit still depends only weakly on the number of modules, i.e., system size.
This means that sweeping $\gamma$ can be used to probe communities of different sizes, and the suitable range
of $\gamma$ values is practically independent of the system size.

These considerations show that
the resolution of the weighted RB method does not necessarily decrease when dense networks grow in size, unlike for sparse networks. 
However, for practical purposes, issues such as the distribution of weights both within and between the blocks
is expected to affect the actual resolution, and the above examples should be viewed as illustrative only.

\section{Example application: modules in a stock correlation network}

As a real-world example, we apply the weighted RB method to a correlation-based network of stock return time series.
Networks of this type are of special interest as the correlations between asset returns are the main input in the classical and still widely used Markowitz portfolio optimization theory \cite{RefWorks:181}. Correlations of stock returns were first studied from the network point of view by Mantegna \cite{RefWorks:183}, who defined a correlation-based metric and was consequently able to identify modules that make sense also from the economic point of view by using the \textit{maximal spanning tree}. This work has been extended by Bonanno \emph{et al.}~\cite{RefWorks:184, RefWorks:185, RefWorks:186} and Onnela \emph{et al.}~\cite{RefWorks:187, RefWorks:188}, with the overall conclusion that there is cluster structure which corresponds well to economic sectors. Recently, the structure of correlation-based stock interaction networks has also been studied with the weighted version of the clique percolation method \cite{RefWorks:173} and by spectral and thresholding analyses \cite{RefWorks:190,RefWorks:192,RefWorks:189,RefWorks:193}.

To construct our network, we use a data set consisting of the daily closing prices of $N=116$ NYSE-traded stocks from the time period from 13-Jan-1997 to 29-Jan-2000\footnote{The length of the time series is 1000 trading days}. We estimate the equal time correlation matrix of logarithmic returns by
\begin{equation} \label{eq:corr}
C_{ij}=\frac{\langle \mathbf{r}_{i} \mathbf{r}_{j}
  \rangle -\langle \mathbf{r}_{i} \rangle \langle \mathbf{r}_{j} \rangle }
  {\sqrt{[\langle {\mathbf{r}_{i}}^{2} \rangle
      -\langle \mathbf{r}_{i}\rangle ^{2}][\langle {\mathbf{r}_{j}}^{2} 
       \rangle -\langle \mathbf{r}_{j} \rangle
      ^{2}]}}, 
\end{equation}
where $\mathbf{r_i}$ is a vector containing the logarithmic returns of stock $i$. Since there is a small number of elements of $\mathbf{C}$ which are slightly negative, we define the weights of our network by
\begin{equation} \label{eq:CW}
W_{ij} = |C_{ij}| - \delta_{ij},
\end{equation}
which can be justified by interpreting the absolute values of correlations as measures of interaction 
strength without considering whether the interaction is positive or negative. 

Here, we take a multiresolution approach to the problem of detecting modules in the above matrix,  and  sweep the value of $\gamma$ to obtain the modules of $W$ at multiple levels of resolution. For each value of $\gamma$, we assign nodes into modules such that the energy (\ref{eq:FM1}) is minimized. Evidently, exploring all possible configurations is computationally impossible, so that some approximative method has to be employed. The choice of method naturally depends on the system size, and for very large systems, greedy optimization methods \cite{Blondel:2008kx,Ronhovde:2008vn} which directly look for local minima might be the only solution. For our case, the system is not very large, and we have chosen the simulated annealing approach, using single-spin flips as well as block flipping as the elementary Monte Carlo operations. It should be noted, however, that it cannot be guaranteed that the obtained energy minimum is a global one. For the RB method, there is no way around this problem.

\begin{figure}[tb]
\begin{center}
\subfigure {\includegraphics[width=0.45\textwidth]{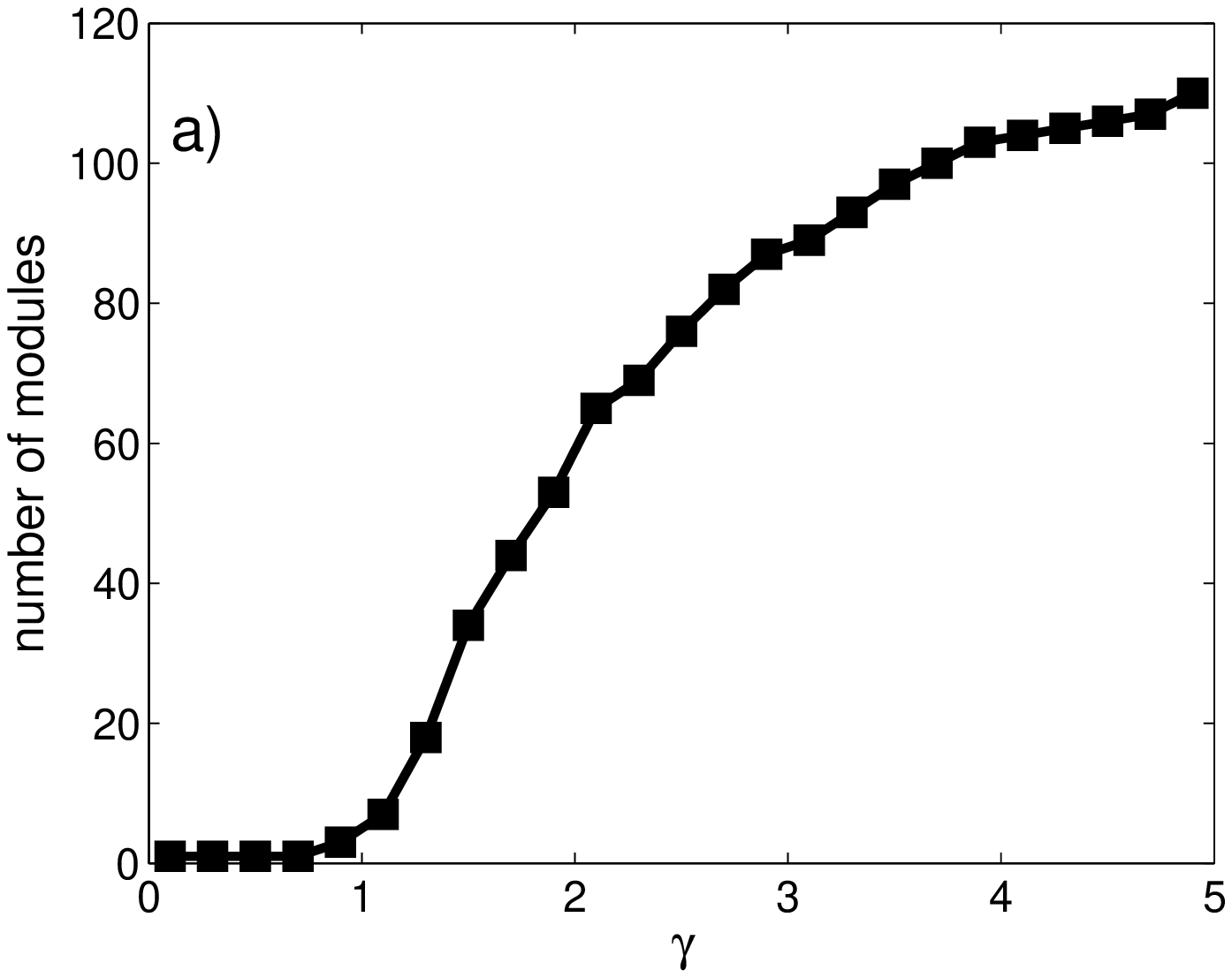}}
\subfigure {\includegraphics[width=0.45\textwidth]{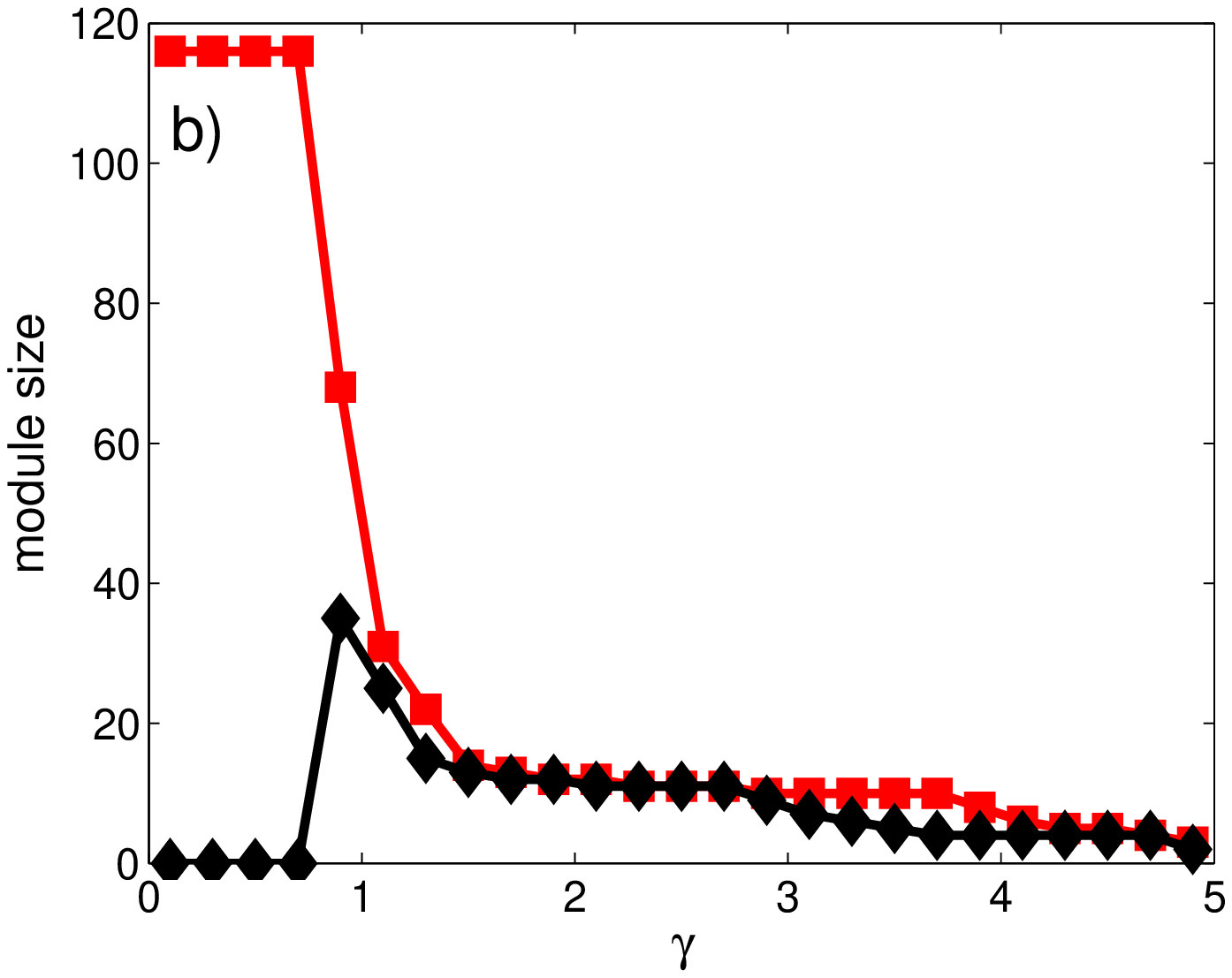}}
\end{center}
\caption{The number of modules (a) and the sizes of the two largest modules (b) as a function of $\gamma$.}
\label{fig:comp_number_size}
\end{figure}

First, we have investigated the number of modules as a function of $\gamma$ (see Fig.~\ref{fig:comp_number_size}a). For $\gamma \lessapprox 0.8$, all nodes are assigned to a single module. When $\gamma$ is further increased, the number of modules starts to rapidly increase, until finally each module corresponds to a single node. It is worth noting that no plateaus are seen in the graph, except for the trivial case of $\gamma \lessapprox 0.8$. In Ref.~\cite{RefWorks:204}, using a related multiresolution method, such plateaus were shown to exist for test-case networks,
corresponding to built-in hierarchical modules. Plateaus would hence yield "natural" choices of the tuning parameter. 
Their absence in Fig.~\ref{fig:comp_number_size}a) means that there is no range of $\gamma$, which would correspond to a stable module configuration. However, stability of the number of modules only gives partial insight into the stability of the modular structure. Especially for real-world networks with modules of different sizes and internal weights, 
changes in this number may only reflect e.g.~splitting of small, weak modules, while the strongest modules remain more or less stable when $\gamma$ is increased. This appears to be the case for our stock interaction network. Panel b) of Fig.~\ref{fig:comp_number_size}  depicts the sizes of the two largest modules as a function of $\gamma$. The sizes remain almost constant for an interval of approx.~$\gamma \in [1.4,3]$, and thus
the increase in the module number can be attributed to splitting of smaller modules.

\begin{figure}[t]
\begin{center}
\subfigure {\includegraphics[width=0.95\linewidth]{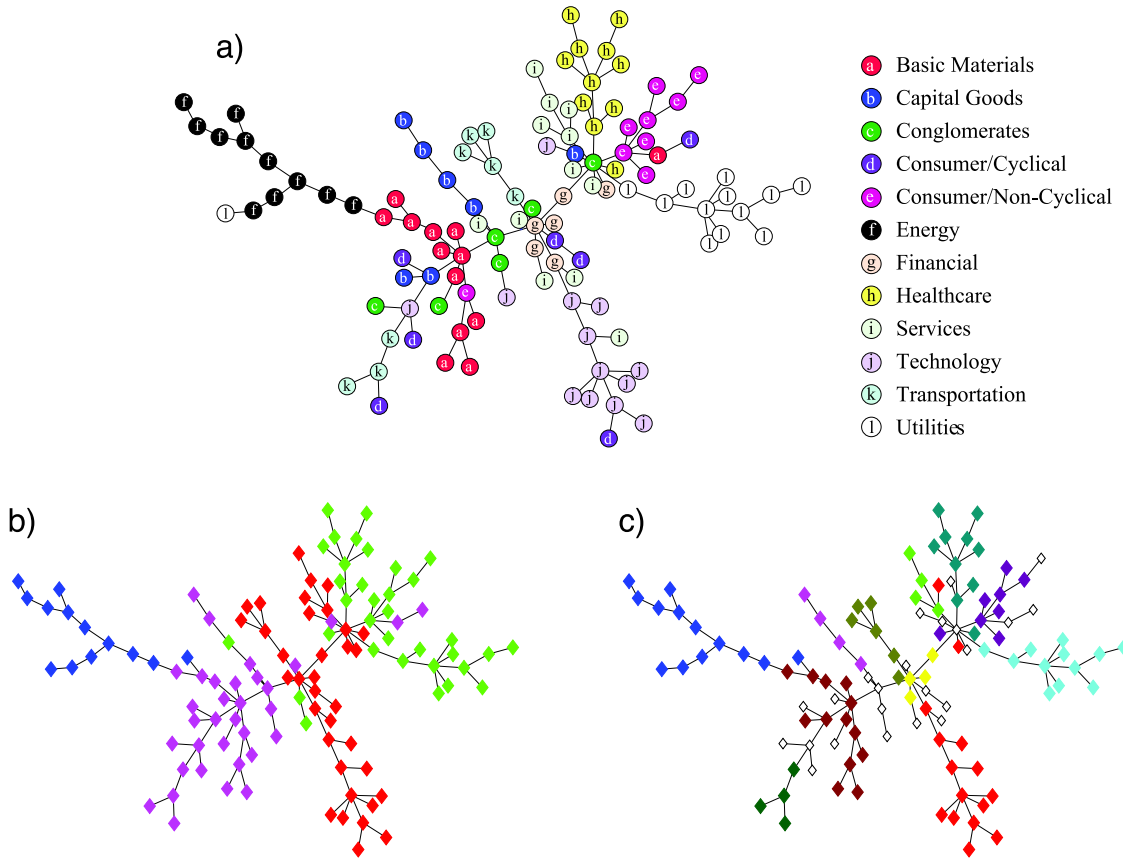}}
\end{center}
\caption{(a) The maximal spanning tree and business sectors according to Forbes \cite{RefWorks:194}. (b) The maximal spanning tree and the modular structure for $\gamma = 1$. Each color corresponds to a module. (c) The maximal spanning tree and the modular structure for $\gamma = 1.4$. Modules of size larger than two are depicted by different colors and the rest of the nodes by empty symbols.}
\label{fig:Potts_mst}
\end{figure}

Next, we turn to the modules themselves. In order to visually compare the detected modules
with known structural features of the investigated correlation matrix, we have utilized the maximal
spanning tree (MST) method. The MST of a network or a matrix is a tree 
connecting all the $N$ nodes with $N-1$ links, such that the sum of the link weights is maximized. 
Earlier, it has been shown that for stock correlation matrices, branches of the MST correspond
well to business sectors or industries for the NYSE~\cite{RefWorks:183,RefWorks:185, RefWorks:186,RefWorks:187,RefWorks:188} as well as FTSE~\cite{Coelho:2007fk}. The typical way to 
categorize stocks into business sectors is to use the Forbes classification \cite{RefWorks:194}.
Panel a) in Figure ~\ref{fig:Potts_mst} displays the MST for the stock network, together with 
the Forbes classification. For comparison, we first set $\gamma=1$ (Fig.~\ref{fig:Potts_mst}b),
and color the nodes according to modules detected by the RB method for the full correlation matrix as above. The value $\gamma = 1$ is of particular interest, as in this case the Hamiltonian of 
Eq.(\ref{eq:RB1}) is equivalent with the weighted version of modularity~\cite{RefWorks:46}. For this value, four modules of sizes 13, 34, 34 and 35 are found. For each module, the majority of member nodes are also connected in the MST, and there is a correspondence between the MST branches and the modules. The smallest module corresponds very well to the Energy sector in the Forbes classification, and the other modules roughly to combinations of different sectors. It should be noted here that the Forbes classification is an external one, \emph{i.e.}, it is not based on empirical observations on stock correlations, and thus some Forbes sectors are also relatively disjoint in the MST of Fig.~\ref{fig:Potts_mst}a). 

Let us now change the resolution of the RB method by moving towards larger values of $\gamma$. Panel c) of Figure \ref{fig:Potts_mst} displays
the modular structure obtained with $\gamma = 1.4$, \emph{i.e.}, at the onset of the "plateau" regime of the two largest module sizes. Only modules of size larger than two are depicted by different colors, while the rest of the nodes are indicated by open symbols. An immediate observation is that the modules correspond remarkably well to the different branches of the MST and very well to the Forbes classification. Increasing $\gamma$ further splits the modules into smaller ones: for $\gamma = 2$ the number of modules is already 58 and thus their average size is only 2. The largest modules, corresponding to the Energy sector and the Electric Utilities industry, are the last ones to break at around $\gamma \approx 3$ and $\gamma \approx 4$, respectively. Interestingly, the Energy module seems to contain a strong submodule of four nodes. This is also seen as a plateau in the graph depicting the size of the second-largest component (Fig.~\ref{fig:comp_number_size}b), which indicates that also large values of $\gamma$ can yield useful information on the modules.

\begin{figure}[t]
\begin{center}
 \subfigure{\includegraphics[width=0.45\textwidth]{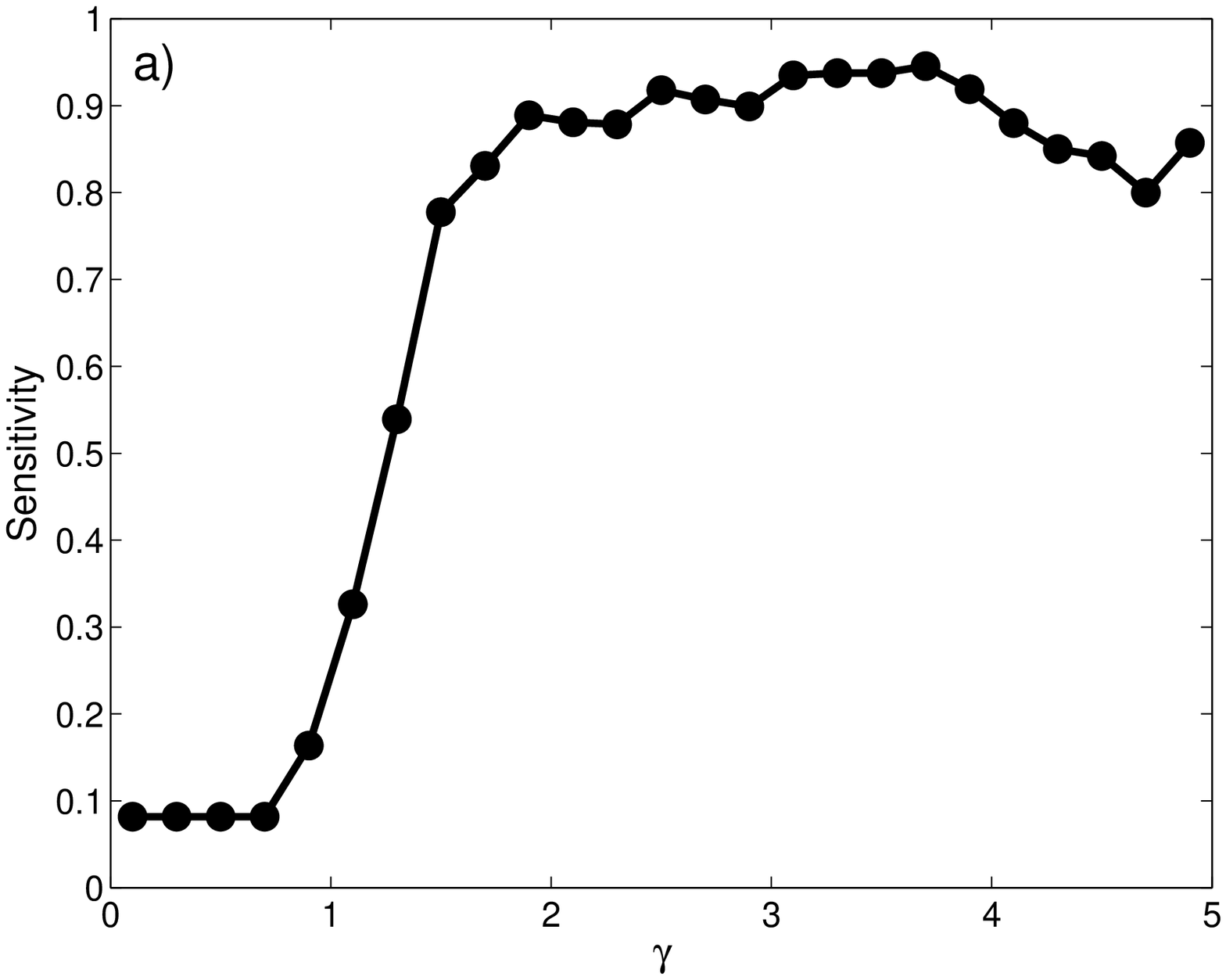} }
 \subfigure {\includegraphics[width=0.45\textwidth]{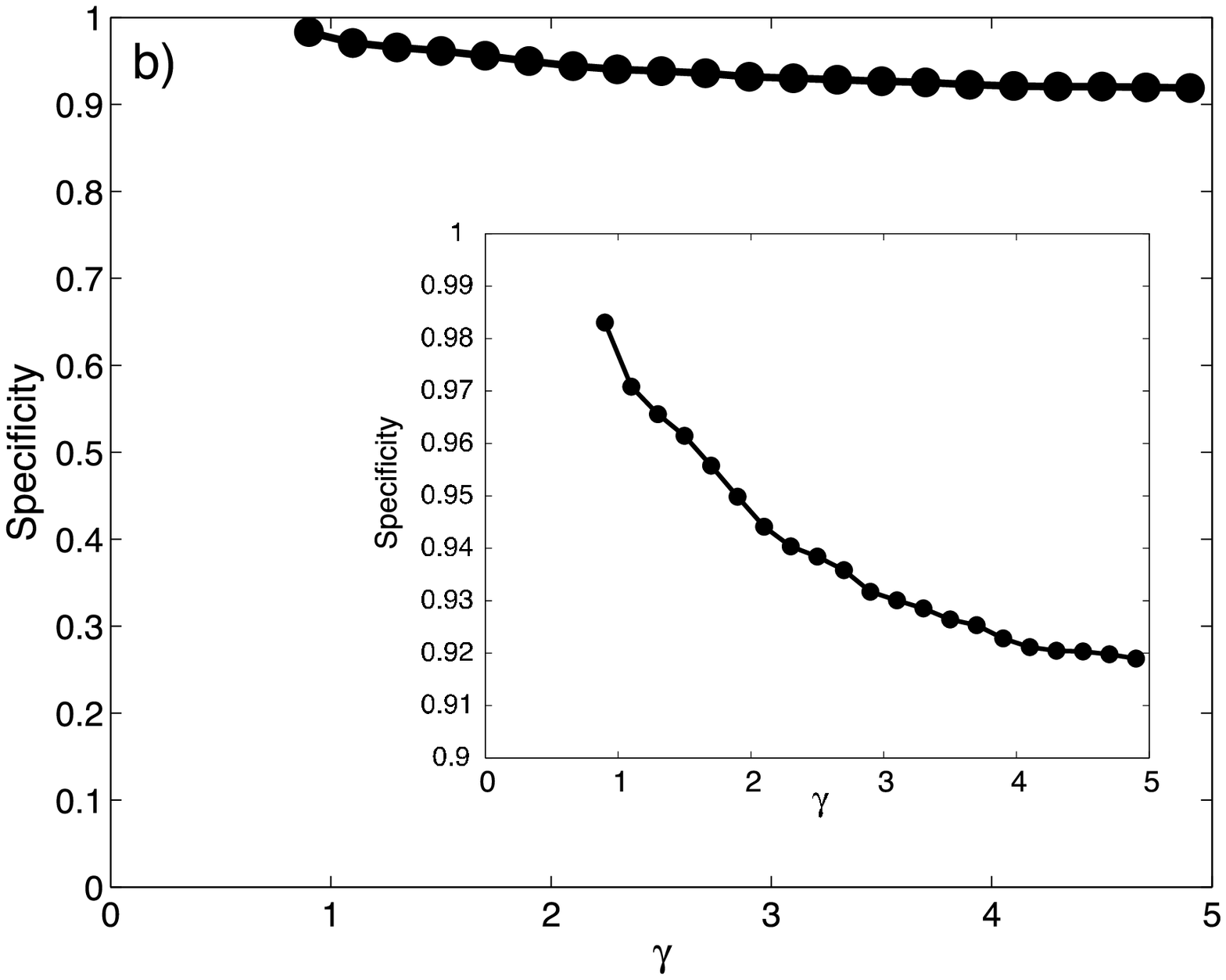}}
\end{center}
\caption{The sensitivity (a) and the specificity (b) of the modular structure with respect to the Forbes classification of business sectors \cite{RefWorks:194} as a function of $\gamma$.
The solid line is a guide to the eye.}
\label{fig:sen_spe}
\end{figure}

Finally, we study the correspondence between the modular structure obtained with the RB method and the Forbes classification to business sectors in a more quantitative way. We use two measures defined in Ref.~\cite{RefWorks:34}: the \emph{sensitivity} defined as the fraction of pairs of nodes classified to the same 
Forbes sector that are assigned to the same module by the RB method  and, correspondingly, \emph{specificity} as the fraction of pairs of nodes 
belonging to different sectors that are assigned to different modules by the RB method. 
Sensitivity and specificity are depicted in Figures \ref{fig:sen_spe}(a) and  \ref{fig:sen_spe}(b), respectively. 
The sensitivity curve shows a sudden increase in the interval $\gamma \in [0.8, 1.8]$. The reason for its low initial value is the assignment of all nodes to a single module, as discussed above, and the 
increase corresponds to 
modules splitting into smaller units which correspond well to the Forbes classification. The high value of sensitivity for large $\gamma$ means that the relatively small modules given by the RB method are proper subsets of the Forbes business sectors.
The specificity curve shows a decreasing trend, but its values still remain relatively high. 
This trend is explained by an increasing number of small modules (including modules consisting of one node only), such that nodes which belong to a common sector appear in different modules.
Overall, the above results indicate that the modular structure detected by the weighted RB method
corresponds well to the Forbes classification for a wide range of $\gamma$, and the small modules
obtained at large $\gamma$ seem to be valid submodules of larger ones.

\begin{figure}[t]
\begin{center}
  \subfigure{\includegraphics[width=0.4\linewidth]{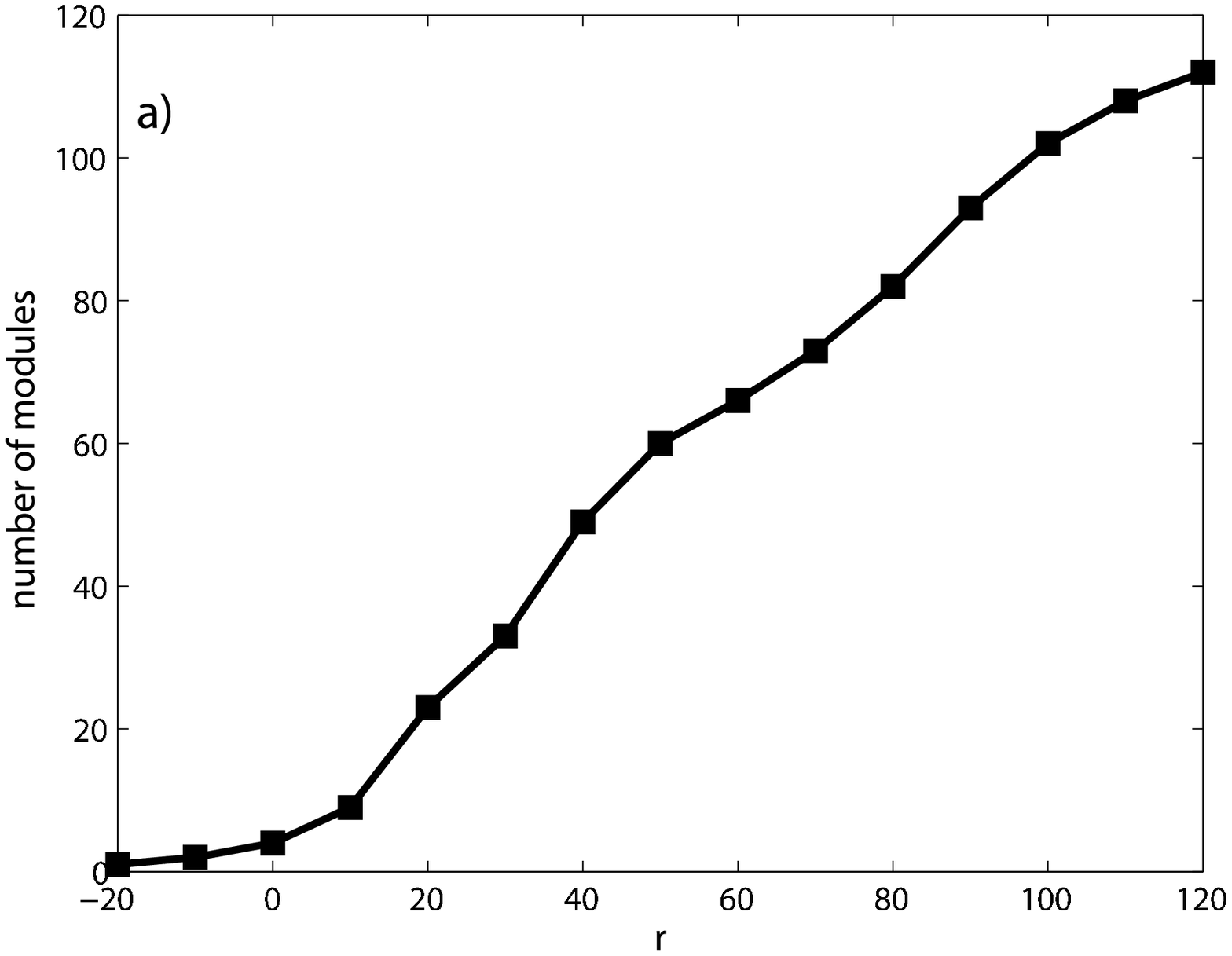} }
  \subfigure{\includegraphics[width=0.4\linewidth]{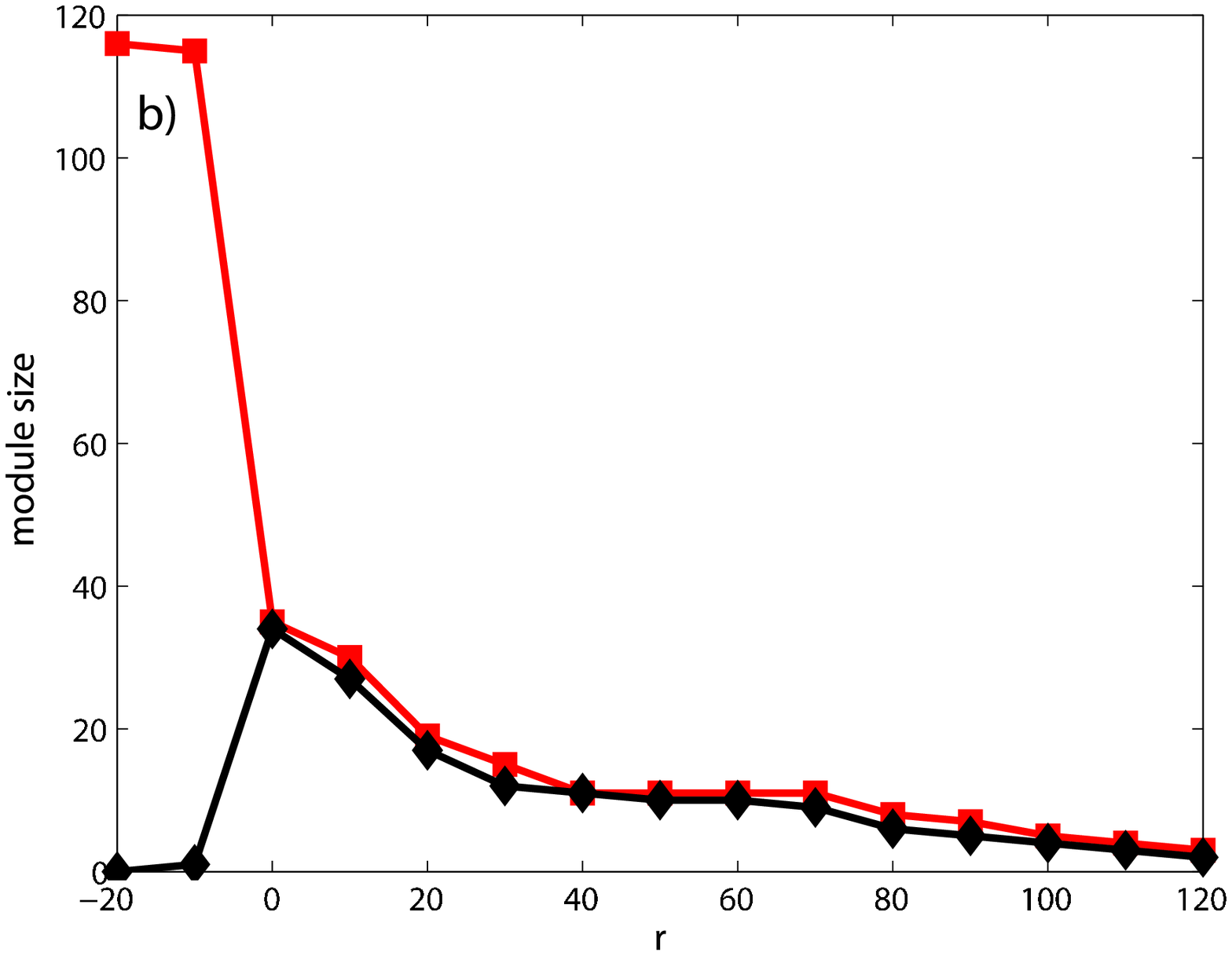}}
  \subfigure{\includegraphics[width=0.4\linewidth]{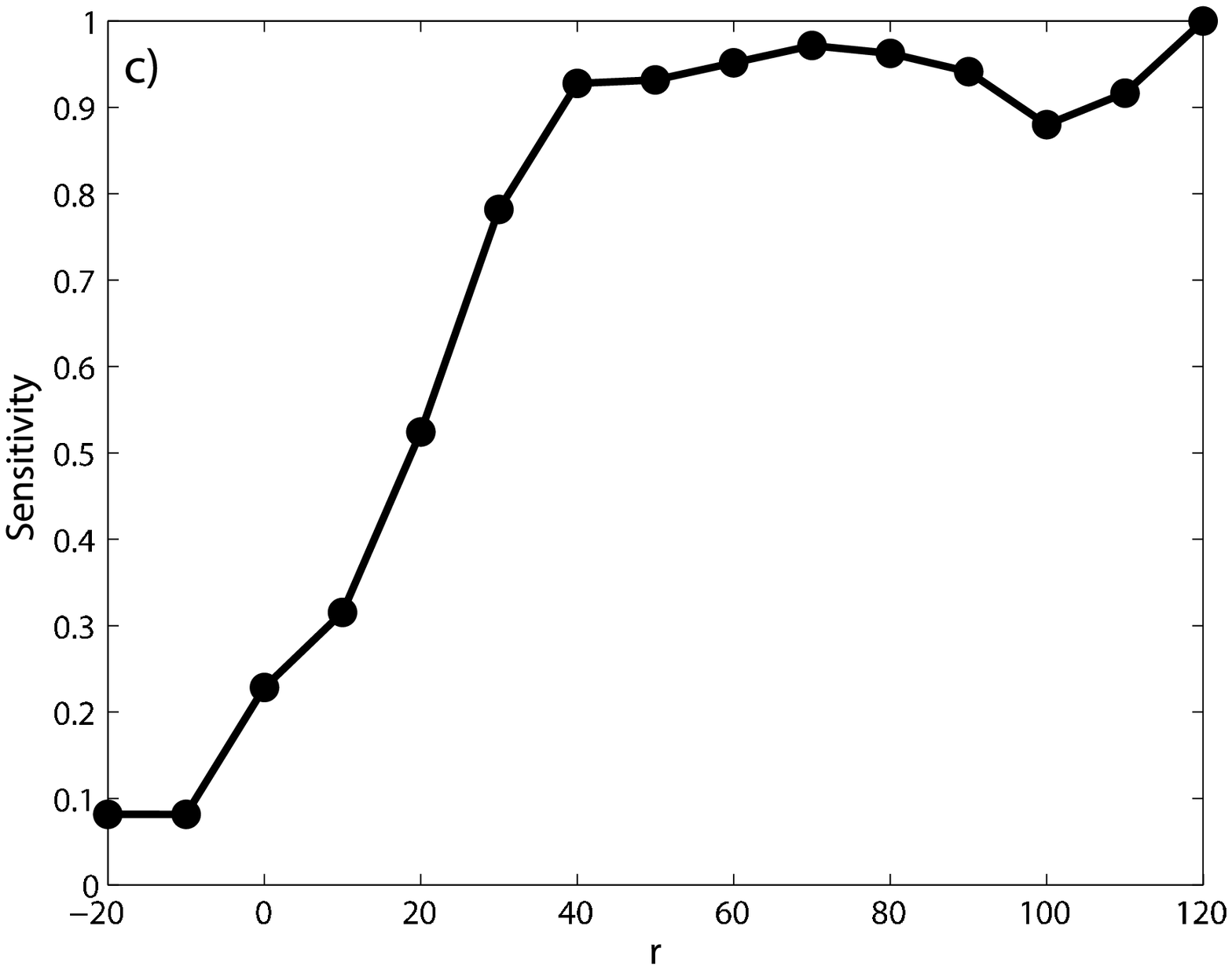} }
  \subfigure{\includegraphics[width=0.4\linewidth]{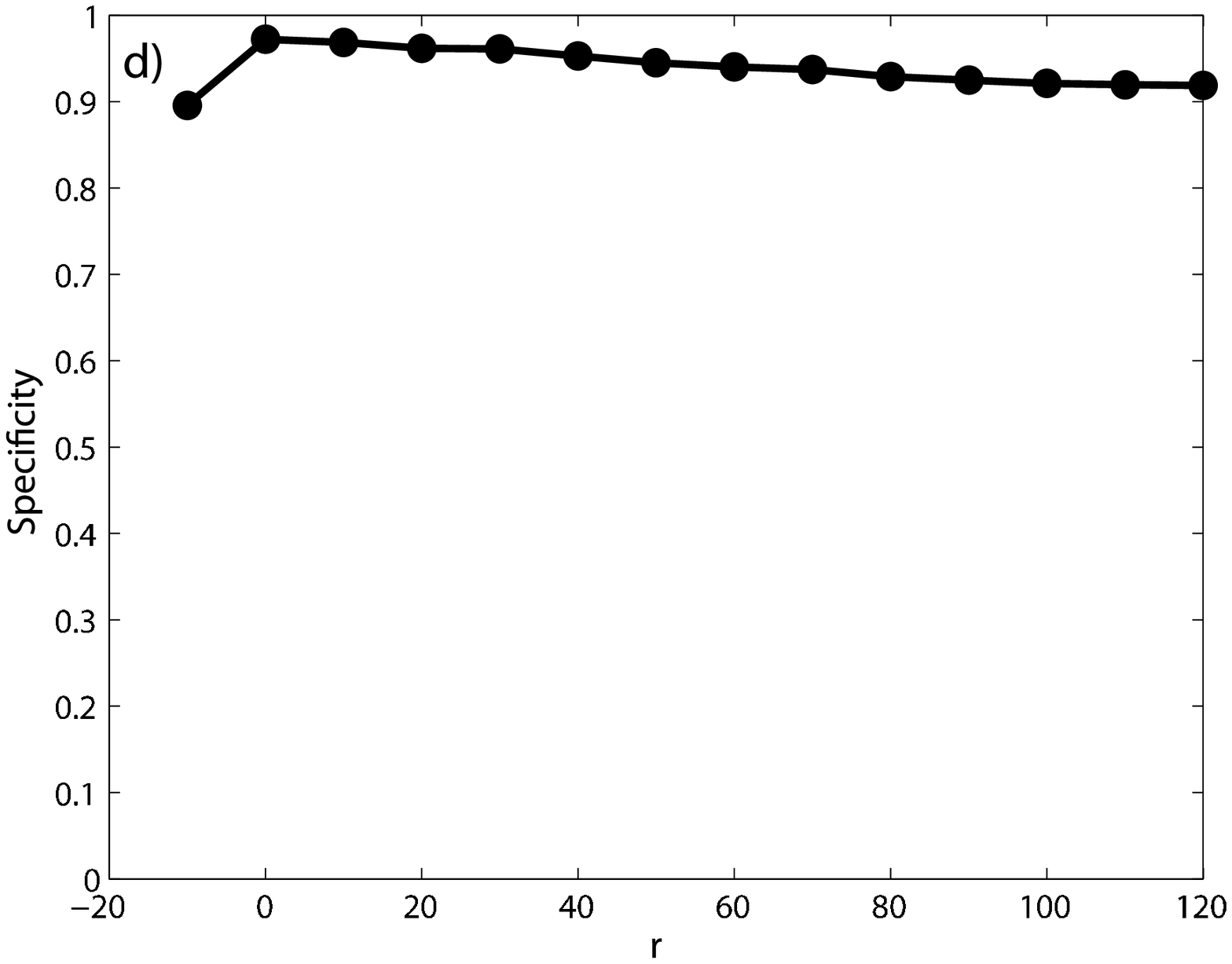}}
\end{center}
\caption{The number of modules (a), the sizes of the two largest  
modules (b), the sensitivity (c) and the specificity (d) as functions  
of $r$ with the AFG method. The solid line is a guide to the eye.}
\label{fig:AFG}
\end{figure}

For comparison, we have also carried out the above analysis using the recently introduced  
weighted multiresolution method by Arenas \textit{et al.}  
\cite{RefWorks:204}. This method resembles the Potts approach; however, the tuning parameter  
$\gamma$ is replaced by the parameter $r$, which can be interpreted as representing
 the weight of a  
self-link added to each node. The number of modules, the sizes of the  
two largest modules, the sensitivity and the specificity as functions  
of the tuning parameter $r$ are depicted in Fig. \ref{fig:AFG}. Comparison with Figs.  
\ref{fig:comp_number_size} and \ref{fig:sen_spe}, in which the same  
results for the RB method are shown, suggests that for the correlation
matrix analyzed here, both the AFG and RB  
methods behave in a very similar manner.

\section{Conclusions}

Here we have presented, analyzed, and applied a weighted version of the $q$-state Potts model approach by Reichardt and Bornholdt~\cite{RefWorks:34}, introducing a well-motivated null model for expected weights within modules. Our target has been to investigate the modular structure of dense weighted networks such that instead of the topology, the link weights determine the modules. In contrast to conventional approaches, where weights considered insignificant are filtered out, our target has been to utilize all information contained in the weight matrix. The weighted RB model fulfills this criterion, as it can equally well be applied to sparse and dense networks. In addition, it contains a parameter that allows tuning its resolution, which is useful for studies of nested community structures. Analysis of the resolution limit of the method has shown that for simple example cases, dense modular networks behave differently from sparse ones as the resolution is only weakly dependent on the network size. As a practical application, we have used the method in analysis of the modular structure of a stock correlation matrix. Our results indicate that by varying the tuning parameter value, the method is able to detect modules which correspond to relevant business sectors, as well as substructure inside these modules. Thus it turns out that the weighted Potts method provides a feasible approach to community detection in dense networks.

\section*{Acknowledgments}
Support by the Academy of Finland, the Finnish Center of Excellence Program 2006-2011, Project No.~213470 is acknowledged. J.K. is partly supported by the GETA graduate school. JS acknowledges support by the European Commission NEST Pathfinder initiative on Complexity through project EDEN (Contract 043251).

\section*{References}
\bibliographystyle{iopart-num}
\bibliography{Heimo1806}

\end{document}